\documentclass[reqno,twoside]{article}
\usepackage{fullpage}

\usepackage{amsmath}
\usepackage{amsfonts}
\usepackage{amssymb}
\usepackage{graphicx}
\usepackage{empheq}
\usepackage{subfigure}
\usepackage{tensor} 
\usepackage{colonequals}
\usepackage{cite}

\usepackage{color}

\usepackage{fancyhdr}
\usepackage[
  pdfview={XYZ null null 1},
  pdfstartview={XYZ null null 1},
  colorlinks,
  linkcolor=blue,
  urlcolor=blue,
  citecolor=blue,
  plainpages=false,
  pdfpagelabels,
]{hyperref}
\hypersetup{
pdftitle={On the spectrum of field quadratures for a finite number of photons}, 
pdfauthor={E Pisanty and E Nahmad-Achar}
}
\usepackage{doi}
\usepackage{url}
\newcommand{\RR}{\ensuremath{\mathbb{R}}} 
\DeclareMathOperator{\Span}{span} 
\DeclareMathOperator{\Tr}{Tr}
\newcommand{\rmd}{\ensuremath{\textrm{d}}}
\newcommand{\Or}{\ensuremath{O}}
\newcommand{\hxi}{\ensuremath{\hat{\xi}}}
\newcommand{\hpi}{\ensuremath{\hat{\pi}}}
\newcommand{\hPi}{\ensuremath{\hat{\Pi}}}
\newcommand{\ha}{\ensuremath{\hat{a}}}
\newcommand{\had}{\ensuremath{\hat{a}^\dagger{}}}
\newcommand{\hn}{\ensuremath{\hat{n}}}
\newcommand{\hil}{\ensuremath{\mathcal{H}}}
\newcommand{\bra}[1]{\ensuremath{\left\langle#1\right|}}
\newcommand{\ket}[1]{\ensuremath{\left|#1\right\rangle}}
\newcommand{\bracket}[2]{\ensuremath{\left\langle#1 \vphantom{#2}\middle|  #2 \vphantom{#1}\right\rangle}}
\newcommand{\matrixel}[3]{\ensuremath{\left\langle #1 \vphantom{#2#3} \middle| #2 \middle| #3 \vphantom{#1#2} \right\rangle}}
\def\urltilda{\kern -.15em\lower .7ex\hbox{\~{}}\kern .04em}

\newcommand{\changed}[1]{{#1}}
\begin{document}%

\title{\Large \bf On the spectrum of field quadratures\\ for a finite number of photons}
\author{\rm E Pisanty, E Nahmad-Achar}
\maketitle

\begin{abstract}
The spectrum and eigenstates of any field quadrature operator restricted to a finite number $N$ of photons are studied, in terms of the Hermite polynomials. By (naturally) defining \textit{approximate} eigenstates, which represent highly localized wavefunctions with up to $N$ photons, one can arrive at an appropriate notion of limit for the spectrum of the quadrature as $N$ goes to infinity, in the sense that the limit coincides with the spectrum of the infinite-dimensional quadrature operator. In particular, this notion allows the spectra of truncated phase operators to tend to the  complete unit circle, \changed{as one would expect.} A regular structure for the zeros of the Christoffel-Darboux kernel is also shown.
\end{abstract}


\section{Introduction}

The optical phase within quantum optics has posed a problem ever since Dirac \cite{Dirac} first formulated an approach to the subject based on Hamilton's equations for the energy and phase as conjugate variables, and using a hermitian operator for the phase which was later proved inconsistent. Several approaches to this problem have included the use of \changed{unphysical} negative number states \cite{negativenumberstates}, nonunitary phase operators \cite{SG,CN}, and the truncation of the Hilbert space to obtain unitary operators.

This last scheme, due to Pegg and Barnett \changed{\cite{NietoReview,PBreview,PBbook}}, consists of restricting the full Hilbert space \hil{} to a subspace containing at most $N$ photons, $\hil^N=\Span\{\ket{0},\ldots,\ket{N}\}$, through the action of the projector $\hPi_N=\sum_{n=0}^N\ket{n}\bra{n}$. Thus a polar decomposition is found for the restricted annihilation operator $\ha_N=\hPi_N\ha\hPi_N$, and the unitary exponential phase operator is used to make physical predictions (in particular those involving its spectrum and eigenvectors) and finally the limit $N\rightarrow\infty$ is taken. 

This approach is valuable in that it permits, as opposed to other alternatives, the construction of a proper Hermitian phase operator, which, however, can be seen as problematic in that it depends on a limiting procedure that \changed{some authors have found unacceptable \cite{criticas}}. After the publication of Pegg and Barnett's work, a number of articles have appeared which treat a number of problems related to the truncation of the dimension of the original Hilbert space, searching, in particular, for coherent states \cite[and references therein]{Roy}. The relationship of the \changed{full space with the} truncated space, and the commutation relations possible in it, has also been studied in \changed{\cite{PBlimitingprocedures,PVBcanonicalconjugation, Hammel, Santhanam, Gudder,delatorre, cotfas,JagannathanI}.}

In this contribution we describe, within this formalism, the spectrum and eigen\-vectors of the position quadrature operator $\hat\xi_N$ for a finite number of photons (and thus for any arbitrary quadrature), making extensive use of Hermite polynomials \cite{Buchdahl, Figurny}. (These polynomials' connections with the harmonic oscillator system are \changed{deep}, and have been studied e.g. in \cite{Manko1, Manko2}.) The normalization of the eigenstates is resolved using the Christoffel-Darboux kernel, and two possible normalizations are given (Section~\ref{quadspectra}).

We then show that \textit{approximate} eigenstates $\ket{\lambda}_N$ can be naturally defined, similarly to \cite{Buchdahl}, for all real (pseudo)-eigenvalues $\lambda$ (Section~\ref{approxev}), and we define and study a quantitative measure, $d_N(\lambda) = ||(\hxi_N-\lambda)\ket{\lambda}_N ||^2$, for the exactness of this approximation. We give suitable approximations for this function and find that its \changed{analytical continuation} into the complex plane has poles on the zeros of the Christoffel-Darboux kernel; we show numerical evidence for a regular structure of the latter.

By studying the zeros of $d_N(\lambda)$, which give the exact eigenstates, one can define an appropriate notion of limit for the spectrum of $\hxi_N$, in the sense that it tends to the full spectrum of the infinite-dimensional position operator $\hxi$ as $N\rightarrow\infty$. We also show that with this notion of limit the finite spectra of the truncated phase operators of Pegg and Barnett (which cover $N+1$ equally spaced points on the unit circle) tend to the complete unit circle -- which is what one would expect from a phase operator --, in contrast to their assertion \changed{in an early paper} \cite{PBunitaryphase} that in the $N\rightarrow\infty$ limit the phase operator should have only a countable number of eigenvalues.

A measure of how good an approximation a given vector is to an eigenvector of the position operator is also given by the spread of its wavefunction. This is studied in Section~\ref{WFsaMs}, where we show that the wavefunctions are highly localized at the pseudo-eigenvalues. We also calculate and approximate the expectation value and dispersion of $\hxi_N$ and $\hxi$. We finish in Section~\ref{final} with a few remarks on the Jacobi-matrix representation of the Hermite polynomials.

This work is presented partly because of its own intrinsic interest, and partly in the hope that it will help to understand better the limiting procedure in Pegg and Barnett's formalism. There one tries to understand a hard-to-define operator \changed{by} approximating it with a sequence of finite-dimensional operators; here we have studied the analogous limiting procedure for the better-understood position operator. \changed{More generally, our results give some indication of features to expect when restricting physical operators with continuous spectra to finite-dimensional subspaces.} 

\changed{Additionally, there are pedagogical applications to the development of finite-dimensional quantum mechanics through the discretization of a particle's position. In that context, the constructions in this paper form a more natural basis than that obtained by taking equally-spaced points.}

The results \changed{presented} here constitute an extension to the work in \cite{tesis}.

\section{Quadrature spectra}
\label{quadspectra}

Since an arbitrary quadrature $\hxi_\beta=\frac{1}{\sqrt{2}}\left(e^{-i\beta} \ha + e^{i\beta}\had\right)$ is unitarily similar to the position quadrature, through the operator $\hat{U}(\beta)=e^{i\beta\hn}$, it suffices to consider only the position quadrature. Thus, we consider the restricted quadrature \changed{$\hxi_N=\frac{1}{\sqrt{2}}\left(\ha_N+\ha_N^\dagger\right)$}, given in the number basis by the matrix
\begin{equation}
\hxi_N=\frac{1}{\sqrt{2}}\left(\begin{array}{cccccc}
0 &     1    &     0    & \cdots & 0 & 0 \\
1 &     0    & \sqrt{2} & \cdots & 0 & 0 \\
0 & \sqrt{2} &     0    & \cdots & 0 & 0 \\
\vdots & \vdots & \vdots & \ddots & \vdots & \vdots \\
0 &     0    &     0    & \cdots &     0    & \sqrt{N} \\
0 &     0    &     0    & \cdots & \sqrt{N} &    0 
\end{array}\right).
\label{matrizhxiN}
\end{equation}
\changed{This operator is the truncation, through $\hxi_N=\hPi_N\hxi\hPi_N$, of the infinite-dimensional position operator $\hxi{}$, whose domain is $D(\hxi)=\{\ket{\psi}\in\hil:\hxi\ket{\psi}\in\hil\}$, and whose spectrum is continuous and consists of all real numbers, with an infinite-norm eigenbasis $\ket{\xi}$ normalized to $\bracket{\xi}{\xi'}=\delta(\xi-\xi')$ such that $\hxi\ket{\xi}=\xi\ket{\xi}$.}

The characteristic polynomial for $\hxi_N$ (modulo a sign), $\det(\hxi+\lambda)$, can be expanded by minors along the last row and then along the last column to give a recurrence relation,
\begin{equation}
\det(\hxi_{N+1}+\lambda)=\frac{2\lambda}{2}\det(\hxi_N+\lambda)-\frac{2N}{2^2}\det(\hxi_{N-1}+\lambda).
\label{recurrenciapolcar}
\end{equation}
This, together with the initial polynomials $\det(\hxi_0+\lambda)=\lambda$ and $\det(\hxi_1+\lambda)=\frac{1}{2}\left(2\lambda^2 -1\right)$, for $N=0$ and $N=1$ respectively, is sufficient to determine the polynomials for all $N$.

This can be used to relate them to the standard families of orthogonal polynomials. Since the Hermite polynomials (as defined in \cite{Stegun}) follow the recurrence relation
\begin{equation}
H_{n+1}(x)=2x H_n(x)-2n H_{n-1}(x),
\label{hermiterecurrence}
\end{equation}
with initial values $H_1(x)=2x$ and $H_2(x)=4x^2-2$, one can easily prove that the characteristic polynomials are given by
\begin{equation}
\det(\hxi_N+\lambda)=\frac{1}{2^{N+1}}H_{N+1}(\lambda).
\label{polcarhermite}
\end{equation}

In particular, this means that the eigenvalues of $\hxi_N$ are exactly the roots of the $(N+1)$\textsuperscript{th} Hermite polynomial. Since there is no exact formula for these roots above the \changed{9th} order (\changed{until} which the parity of the polynomials allows exact solutions in terms of radicals), one might think at first that the diagonalization of $\hxi_N$ is an unassailable problem.

However, the solution can indeed be found by leaving the eigenvalue $\lambda$ alone and reducing the system $\left(\hxi_N-\lambda\right)\ket{\lambda}_N=0$ to upper triangular form using Gaussian elimination; after this, the system is bidiagonal with coefficients in terms of the Hermite polynomials. (Here we have switched from the combination $\hxi_N+\lambda$ to $\hxi_N-\lambda$ to simplify the algebra, but no results are affected since as the Hermite polynomials have definite parity the zeros are unchanged.) The solution is then found to be
\begin{equation}
\ket{\lambda}_N=c_N(\lambda)\sum_{n=0}^N \frac{H_n(\lambda)}{\sqrt{2^n n!}}\ket{n},
\label{solucion}
\end{equation}
in which $c_N(\lambda)$ is a (nonzero) normalization constant which will be discussed later. 

Independently of how it was obtained, expression \eqref{solucion} can easily be seen to be a solution to the system $\left(\hxi_N-\lambda\right)\ket{\lambda}_N=0$ by expressing the latter as the equations
\begin{subequations}
\label{sistema}
\begin{empheq}[left=\empheqlbrace]{alignat=6}
&-\lambda\bracket{0}{\lambda}_N + & \frac{1}{\sqrt{2}}&\bracket{1}{\lambda}_N&=0,&\label{sistema.cero}\\
\sqrt{\frac{n}{2}}\bracket{n-1}{\lambda}_N&-\lambda\bracket{n}{\lambda}_N+ & \sqrt{\frac{n+1}{2}}&\bracket{n+1}{\lambda}_N&=0,& \qquad 0<n<N, \label{sistema.medio}\\
\sqrt{\frac{N}{2}}\bracket{N-1}{\lambda}_N&-\lambda\bracket{N}{\lambda}_N & & &=0.& \label{sistema.fin}
\end{empheq}
\end{subequations}

The solution \eqref{solucion} satisfies \eqref{sistema.cero} trivially, and \eqref{sistema.medio} can be seen to be equivalent to the Hermite polynomials' three-term recurrence relation \eqref{hermiterecurrence}. It is only \eqref{sistema.fin} which depends on $\lambda$ being an eigenvalue for \hxi{}, since from the Hermite recurrence relation
\[
\sqrt{\frac{N}{2}}\bracket{N-1}{\lambda}_N-\lambda\bracket{N}{\lambda}_N=-\frac{1}{2}\frac{c_N(\lambda)}{\sqrt{2^N N!}}H_{N+1}(\lambda).
\]

Thus, for \textit{any} $\lambda$ the vector $\ket{\lambda}_N$ differs only in a single component from being an eigenvector of $\hxi_N$:
\begin{equation}
\left(\hxi_N-\lambda\right)\ket{\lambda}_N = -\frac{1}{2}\frac{c_N(\lambda)}{\sqrt{2^N N!}}H_{N+1}(\lambda)\ket{N}.
\label{casieigenvector}
\end{equation}

Since the $(N+1)$\textsuperscript{th} Hermite polynomial has $N+1$ zeros, this construction gives as many eigenvectors as the dimension of $\hil^N$ and therefore solves the diagonalization problem. The normalization, however, is still to be fixed. It is natural to ask that the eigenvectors have unit norm: that is, to require that
\begin{equation}
|c_N(\lambda)|^2\sum_{n=0}^N \frac{H_n(\lambda)^2}{2^n n!}={^{}_N}\!\bracket{\lambda}{\lambda}_N=1.
\label{normunitaria}
\end{equation}
Here the summation can be resolved into a much simpler form through the use of the Christoffel-Darboux formula \cite[eq. (22.12.1)]{Stegun} for the Hermite polynomials, 
\begin{equation}
\sum_{n=0}^N\frac{H_n(\lambda)H_n(\mu)}{2^n n!}=\frac{1}{2^{N+1} N!}\frac{H_{N+1}(\lambda)H_N(\mu)-H_N(\lambda)H_{N+1}(\mu)}{\lambda-\mu},
\label{CDdoble}
\end{equation}
which in the limit $\mu\rightarrow\lambda$ gives
\begin{subequations}
\label{CDsimple}
\begin{align}
\sum_{n=0}^N\frac{H_n(\lambda)^2}{2^n n!}&=\frac{H_N(\lambda)H_{N+1}'(\lambda)-H_{N+1}(\lambda)H_N'(\lambda)}{2^{N+1} N!}\label{CDsimple.dif}\\
&=\frac{(N+1)H_N(\lambda)^2- N H_{N+1}(\lambda)H_{N-1}(\lambda)}{2^{N} N!}.\label{CDsimple.alg}
\end{align}
\end{subequations}

The Christoffel-Darboux kernel is of particular importance in this setting, since it is in fact part of the integral kernel of the projection operator $\hPi_N=\sum_{n=0}^N \ket{n}\bra{n}$. This can be seen by calculating its matrix elements in the position representation, \matrixel{\xi}{\hPi_N}{\xi'}, using the fact that the number-basis wavefunctions are given by \cite[eq. $B_V.33$]{CohenTannoudji}
\[
\bracket{\xi}{n}=\frac{1}{\sqrt{2^n n!\sqrt{\pi}}}H_n(\xi)e^{-\xi^2/2}.
\]
One obtains
\begin{align*}
\matrixel{\xi}{\hPi_N}{\xi'}
&= \sum_{n=0}^{N}\bracket{\xi}{n}\bracket{n}{\xi'}
=\sum_{n=0}^{N}\frac{H_n(\xi)H_n(\xi')}{2^n n! \sqrt{\pi}} e^{-(\xi^2+{\xi'}^2)/2} \\
&=\frac{1}{2^{N+1} N! \sqrt{\pi}}\frac{H_{N+1}(\xi)H_N(\xi')-H_N(\xi)H_{N+1}(\xi')}{\xi-\xi'}e^{-(\xi^2+{\xi'}^2)/2},
\end{align*}
which can be used to obtain the projector's action on wavefunctions as
\begin{equation}
\matrixel{\xi}{\hPi_N}{\psi} = \frac{1}{2^{N+1} N! \sqrt{\pi}} e^{-\xi^2/2} \int_{-\infty}^\infty  \frac{H_{N+1}(\xi)H_N(\xi')-H_N(\xi)H_{N+1}(\xi')}{\xi-\xi'}e^{-{\xi'}^2/2} \psi(\xi')\rmd\xi'.
\end{equation}

In this context, the normalization constant $c_N(\lambda)$ is related to the diagonal matrix elements of the projector via
\[
\frac{1}{\sqrt{\pi}|c_N(\xi)|^2}=\sum_{n=0}^{N}\frac{H_n(\xi)^2}{2^n n!\sqrt{\pi}}=e^{\xi^2}\matrixel{\xi}{\hPi_N}{\xi},
\]
where $\lambda$ has been replaced by $\xi$ \changed{in $c_N(\lambda)$}. Additionally, if \eqref{normunitaria} is used as a \textit{definition} for $c_N(\lambda)$, one can use this to obtain the identity
\[
\int_{-\infty}^\infty \frac{e^{-\lambda^2}}{|c_N(\lambda)|^2} \frac{\rmd\lambda}{\sqrt{\pi}} = \Tr(\hPi_N) = N+1.
\]

Of course, the normalization ${^{}_N}\!\bracket{\lambda}{\lambda}_N=1$ is not the only interesting choice. One other important option is to have $c_N(\lambda)$ be independent of $N$; in that case, it is particularly fruitful to have $c_N(\lambda)= \pi^{-1/4} e^{-\lambda^2/2}$, so that the eigenvectors become
\[
\ket{\lambda}_N=\sum_{n=0}^N \frac{H_n(\lambda)}{\sqrt{2^n n!\sqrt{\pi}}}e^{-\lambda^2/2}\ket{n}=\hPi_N\ket{\lambda},
\]
which are the truncation through $\hPi_N$ of the full (infinite-dimensional) position eigen\-vectors with eigenvalue $\lambda$. These truncated eigenvectors have a finite norm which is a monotone function of $N$; in fact, with this normalization, ${^{}_N}\!\bracket{\lambda}{\lambda}_N\rightarrow\infty$ as $N\rightarrow\infty$, which reflects the fact that infinite-dimensional position eigenstates have infinite norm.

\section{Approximate eigenvectors}
\label{approxev}
For each of the $N+1$ zeros $\lambda$ of $H_{N+1}(\lambda)=0$, we have shown the existence of an eigenvector $\ket{\lambda}_N$ of $\hxi_N$, with $\hxi_N\ket{\lambda}_N=\lambda\ket{\lambda}_N$. However, the vectors $\ket{\lambda}_N$ can be defined through \eqref{solucion} even if $\lambda$ is not an eigenvalue of $\hxi_N$. In that case $\ket{\lambda}_N$ is of course no longer an eigenvector, but we have seen that it is ``almost'' one, in the sense that only one component disrupts this behaviour; further, if one is to take the limit $N\rightarrow\infty$ then the \ket{N} component would seem to lose importance.

Thus, we propose to view the vector $\ket{\lambda}_N$ as an \textit{approximate} eigenvector for all $\lambda$, and to judge the exactness of the approximation by the function
\begin{equation}
d_N(\lambda)\colonequals\left|\left(\hxi_N-\lambda\right)\ket{\lambda}_N\right|^2=\frac{|c_N(\lambda)|^2}{2^{N+2} N!}H_{N+1}(\lambda)^2.
\label{defdn}
\end{equation}
\changed{It is important to note that these are neither eigenvectors nor eigenvalues. The index $\lambda$ is perhaps best referred to as a pseudoeigenvalue; while there is a continuum of these, there is of course only a finite number of eigenvalues. Analogues of these states have been briefly considered for angle variables in \cite{PBangle}.}

It is natural to demand the normalization ${^{}_N}\!\bracket{\lambda}{\lambda}_N=1$ so that the length of $\ket{\lambda}_N$ does not affect the limiting behaviour of $d_N(\lambda)$\changed{; one} then has 
\[
d_N(\lambda)=\frac{1}{2}\frac{H_{N+1}(\lambda)^2}{H_N(\lambda)H_{N+1}'(\lambda)-H_{N+1}(\lambda)H_N'(\lambda)}.
\]
\changed{This} can be simplified using the identities 
\[
H_N(\lambda)H_{N+1}'(\lambda)-H_{N+1}(\lambda)H_N'(\lambda)=-H_{N+1}(\lambda)^2 \frac{\rmd}{\rmd\lambda}\left[\frac{H_N(\lambda)}{H_{N+1}(\lambda)} \right]
\] 
and 
\[
\frac{H_N(\lambda)}{H_{N+1}(\lambda)}=\frac{1}{2(N+1)}\frac{H_{N+1}'(\lambda)}{H_{N+1}(\lambda)}=\frac{1}{2(N+1)}\frac{\rmd}{\rmd\lambda}\ln\left| H_{N+1}(\lambda)\right|
\] to the remarkable form
\begin{equation}
d_N(\lambda)
=\frac{-(N+1)}{\frac{\rmd^2}{\rmd\lambda^2}\left[\ln\left|H_{N+1}(\lambda)\right|\right]}.
\label{dnderlog}
\end{equation}

One can also express this as
\begin{equation}
\frac{1}{d_N(\lambda)}
=\frac{\rmd^2}{\rmd\lambda^2}\left[\ln\left(\frac{1}{\sqrt[N+1]{\left|H_{N+1}(\lambda)\right|}}\right)\right],
\label{invdnderlog}
\end{equation}
which immediately gives an asymptotic expression for large $\lambda$, since then $H_{N+1}(\lambda)\approx 2^{N+1} \lambda^{N+1}$. After differentiation, this means that $d_N(\lambda)\approx \lambda^2$ for large $\lambda$, which makes sense: one expects the vector $\ket{\lambda}_N$ to be an increasingly poor approximation for an eigenvector away from the region with eigenvalues.

However, one can say much more about $d_N(\lambda)$: for instance, expression \eqref{dnderlog} does not readily give much information, and it is not very practical for algebraic manipulation \changed{or} numerical evaluation. For such purposes, it is more convenient to use the expression
\begin{equation}
d_N(\lambda)=\frac{1}{4}\frac{H_{N+1}(\lambda)^2}{(N+1)H_N(\lambda)H_{N}(\lambda)-N H_{N+1}(\lambda)H_{N-1}(\lambda)}.
\label{exprdnalgebraica}
\end{equation}

\begin{figure}[htb]
	\centering
		\includegraphics[width=0.9\textwidth]{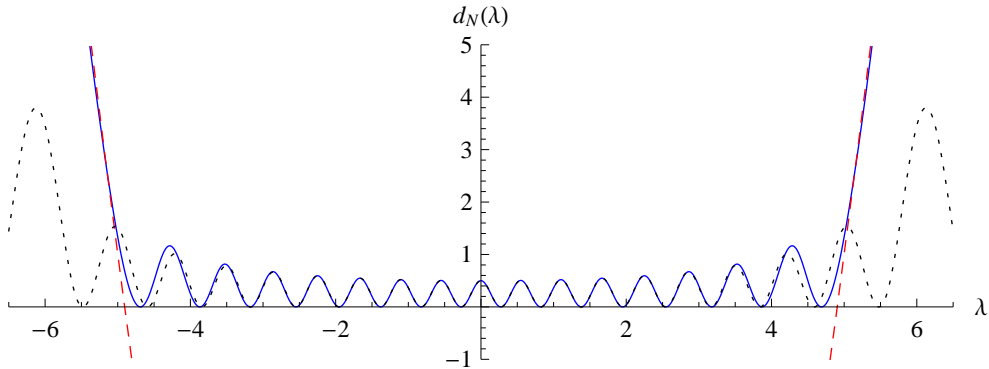}
	\caption{The function $d_N(\lambda)$ and the approximations \eqref{aproxcuad}, up to the constant term (dashed, red), and \eqref{aproxosc}, with the series truncated as they are shown in the text (dotted, black), for $N=15$. }
	\label{figdn}
\end{figure}

The behaviour of $d_N(\lambda)$ splits into two regions: it is oscillatory for small $|\lambda|$ and quadratically growing for large $|\lambda|$. Specifically, one can prove that
\begin{subequations}
\label{aproxes}
\begin{empheq}[left=\empheqlbrace]{align}
d_N(\lambda)&=\lambda^2-\frac{3N}{2}+\frac{N(5-N)}{4\lambda^2}+\cdots \label{aproxcuad}\\
&\text{for} \,\lambda >\sqrt{2N+1},\nonumber \\
d_N(\lambda)&\approx\frac{1}{2}\frac{
1+(-1)^{N+1}\cos\left(2\sqrt{2N+3}\lambda\left(1-\frac{1}{6}\frac{\lambda^2}{2N+3}+\cdots\right)\right)
}{
1+\cos\left(\frac{2}{\sqrt{2N+1}}\lambda\left(1+\frac{1}{12}\frac{\lambda^2}{2N+1}+\cdots\right)\right)
}+\Or (N^{-1}) \label{aproxosc}\\
&\text{for}\,\lambda<\sqrt{2N+1}.\nonumber
\end{empheq}
\end{subequations}
The function $d_N(\lambda)$, along with both of these approximations, is shown in figure \ref{figdn}.

The first of these approximations can easily be expected since $d_N(\lambda)$ is a rational function of $\lambda$; in fact, it is simply a Laurent series, valid \changed{for sufficiently large $\lambda$}. As such, it can be expressed as $d_N(\lambda)=\sum_{k=0}^\infty \alpha_k \lambda^{2-2k}$, where all the coefficients $\alpha_k$ can be found either manually (expressing $H_{N+1}(\lambda)$ as a sum of $\lambda^2\left((N+1)H_N(\lambda)H_{N}(\lambda)-N H_{N+1}(\lambda)H_{N-1}(\lambda)\right)$ and terms of $\Or (\lambda^{2N})$, and so on) or analytically, integrating around a large circle in the complex plane, as
\[
\alpha_k=\frac{1}{2\pi i}\oint\lambda^{2k-3}d_N(\lambda)\rmd\lambda.
\]
This reduces in turn to a sum of residues over the poles of $d_N(\lambda)$, which are exactly the zeros of the Christoffel-Darboux kernel, expression \eqref{CDsimple}. Since the kernel is a sum of squares, it is never zero for real variables (a fact which is often used to prove interpolation theorems for zeros of orthogonal polynomials) so the poles have nonzero imaginary parts. 

\begin{figure}[htbp]
	\centering
		\includegraphics[width=0.9\textwidth]{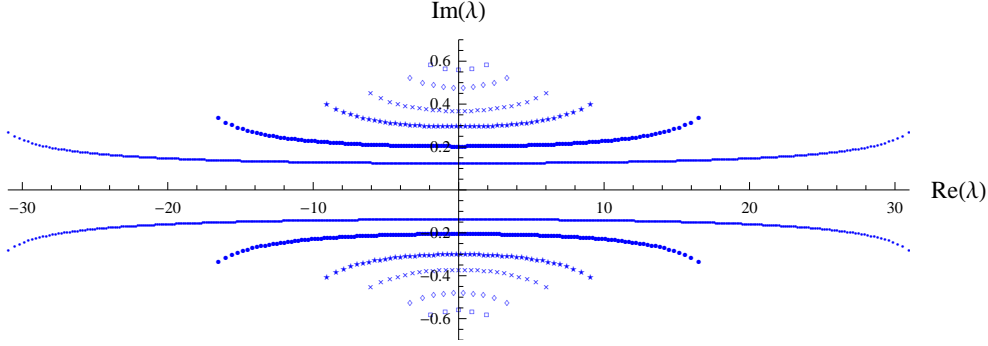}
	\caption{Zeros of the Christoffel-Darboux kernel for the Hermite polynomials, given by $\sum_{n=0}^N H_n(\lambda)^2/2^nn!$, on the complex $\lambda$ plane, for $N=5$, 10, 25, \changed{50, 150 and 500}. Care must be taken since the scales on both axes do not \changed{coincide.}}
	\label{ceros}
\end{figure}

However, despite the extensive attention to the Christoffel-Darboux kernel in the orthogonal-polynomials literature \changed{\cite{Simon}}, \changed{to the best of our knowledge there are no} suitable expressions or approximations for these zeros. It is clear, on the other hand, that these zeros do have some additional structure, which can easily be seen by plotting them: they appear to lie, \changed{with increasingly even spacing}, on two complex-conjugate curves increasingly close to the real axis as $N$ increases. A plot of these zeros for $N=5$, 10, 25, \changed{50, 150 and 500} is shown in figure \ref{ceros}.

The second approximation, \eqref{aproxosc}, correct to order $\Or (N^{-1})$ uniformly for bounded $\lambda$, is obtained from the approximation
\begin{align*}
e^{-x^2/2}H_n(x)&=2^{\frac{n}{2}+\frac{1}{4}}\sqrt{n!}(\pi n)^{-1/4}(\sin(\phi))^{-1/2}\\
&\qquad\times\left[\sin\left(\left(\frac{n}{2}+\frac{1}{4}\right)\left(\sin(2\phi)-2\phi\right)+\frac{3\pi}{4}
\right)+\Or (n^{-1})\right],
\end{align*}
where $x=\sqrt{2n+1}\cos(\phi)$ and $\phi\in[\epsilon,\pi-\epsilon]$, found in \cite[p. 201]{Szego}, by further approximating the argument of the sine function using the Taylor series
\[
\sin(2\phi)-2\phi=-\pi+\frac{4}{\sqrt{2n+1}}x-\frac{4}{(2n+1)^{3/2}}\frac{x^3}{6}+O\left(\frac{x^5}{n^{5/2}}\right)
\]
to obtain 
\begin{align*}
e^{-x^2/2}H_n(x)&\approx \sqrt{\frac{2^n n!\sqrt{2}}{\sqrt{\pi n}}}\left(\frac{2n+1}{2n+1-x^2}\right)^{1/4}  \\
&\qquad\times\cos\left( \sqrt{2n+1} x\left(1-\frac{1}{6}\frac{x^2}{2n+1}+\cdots\right)+n\frac{\pi}{2}\right).
\end{align*}

One consequence is that {}\changed{the spacing of the zeros of $d_N(\lambda)$ (which give the eigenvalues of $\hxi_N$) decreases as $\pi/\sqrt{2N+1}$ for large $N$, and that this spacing is increasingly even as $N$ increases.} Furthermore, the region with zeros covers the increasingly large intervals $-\sqrt{2N+1}\leq\lambda\leq\sqrt{2N+1}$.

One can then show that, given any real number $\lambda$, a positive number $\epsilon$ and an integer $N_0$, there exists an eigenvalue $\lambda_0$ of $\hxi_N$ with $N\geq N_0$ such that $|\lambda-\lambda_0|<\epsilon$. This can in turn be used to define an appropriate notion of the limit of the spectra $\sigma(\hxi_N)$, with the desired property that the spectra tend to the spectrum of the full, infinite-dimensional, position operator \hxi:
\begin{equation}
\lim_{N\rightarrow\infty}\sigma(\hxi_N):=\bigcap_{k=0}^\infty \overline{\bigcup_{N=k}^\infty\sigma(\hxi_N)}=\RR=\sigma(\hxi),
\label{limiteespectro}
\end{equation}
where the overline means topological closure in \RR.

This notion of a limit of a sequence of sets means that the points in the limit set are arbitrarily close to sets of the sequence with arbitrarily large indexes, and it resembles the limit superior \cite[p.16]{Halmos} of a sequence of sets, which omits the topological closure. Additional uses for this definition can be easily found: for example it is in this sense that the hyperboloids $x^2+y^2-z^2=\pm a^2$ ``tend to'' the cone $x^2+y^2=z^2$ as the waist $a$ tends to zero, and that the ellipsoids $x^2+y^2+z^2/c^2=1$, when $c\rightarrow\infty$ and they become increasingly elongated, ``tend to'' the cylinder $x^2+y^2=1$.

Our definition \eqref{limiteespectro} also allows the finite spectra of the truncated phase operators of Pegg and Barnett (which cover $N+1$ equally spaced points on the unit circle) to tend to the complete unit circle, which is what one would expect from a phase operator. This is in contrast to Pegg and Barnett's assertion \changed{\cite{PBunitaryphase}} that in the $N\rightarrow\infty$ limit the phase operator should have only a countable number of eigenvalues: it is clearly dependent on the notion of limit used, which should reproduce in the simpler example of the truncated quadratures $\hxi_N$ the required behaviour.

Another limit that deserves notice is that of $\ket{\lambda}_N$ as $|\lambda|\gg\sqrt{2N+1}$, where the number-basis components in \eqref{solucion} corresponding to \ket{n} with $n<N$, which scale as $2^n\lambda^n$, are drowned out by the \ket{N} contribution. One therefore has, up to a phase, $\ket{\lambda}_N\rightarrow\ket{N}$ as $\lambda\rightarrow\pm\infty$. This limit will become apparent on the graphs of the wavefunctions $\bracket{\xi}{\lambda}_N$ in figure \ref{WFs}.

One further detail of the asymptotic behaviour of $d_N(\lambda)$ is noteworthy, and it relates to the amplitude of the oscillations, which is basically independent of $N$: $d_N(\lambda)$ oscillates sinusoidally from 0 to 1, instead of -- as one might hope -- tending uniformly to zero. This means that for any real $\lambda$, there will always be values $\lambda_1$ arbitrarily close to $\lambda$ that will have $d_N(\lambda_1)=1$, with arbitrarily large $N$.

The fact that $d_N(\lambda_1)=1={^{}_N}\!\bracket{\lambda_1}{\lambda_1}_N$ can be seen as implying that $\ket{\lambda_1}_N$ is as far away as it can be from being an eigenvector, which would disqualify the curve $\lambda\mapsto\ket{\lambda}_N$ in $\hil^N$ from describing approximate eigenvectors. However, this overlooks the fact that $\hxi_N$ has a relatively large (operator) norm, and that one can find states such as \ket{N}, the last number-basis vector, for which $\left\|\left(\hxi- \lambda\right) \ket{N} \right\|^2  = \lambda^2+\frac{N}{2}$. Since \ket{N} represents the wavefunction within $\hil^N$ with arguably the largest position dispersion, it can be seen that the vectors $\ket{\lambda}_N$ are in fact rather close to being position eigenvectors.

\section{Wavefunctions and moments}
\label{WFsaMs}
As we mentioned at the end of last section, one qualitative measure of how good an approximation to an eigenvector of $\hxi_N$ is a given vector is the spread of its wavefunction. In this respect, one might hope that the vectors $\ket{\lambda}_N$ represent wavefunctions highly localized around $\lambda$ with small dispersion.

\begin{figure}[ht!]
\centering
\subfigure[$\lambda=-5$ ]{
  \includegraphics[width=0.4\textwidth]{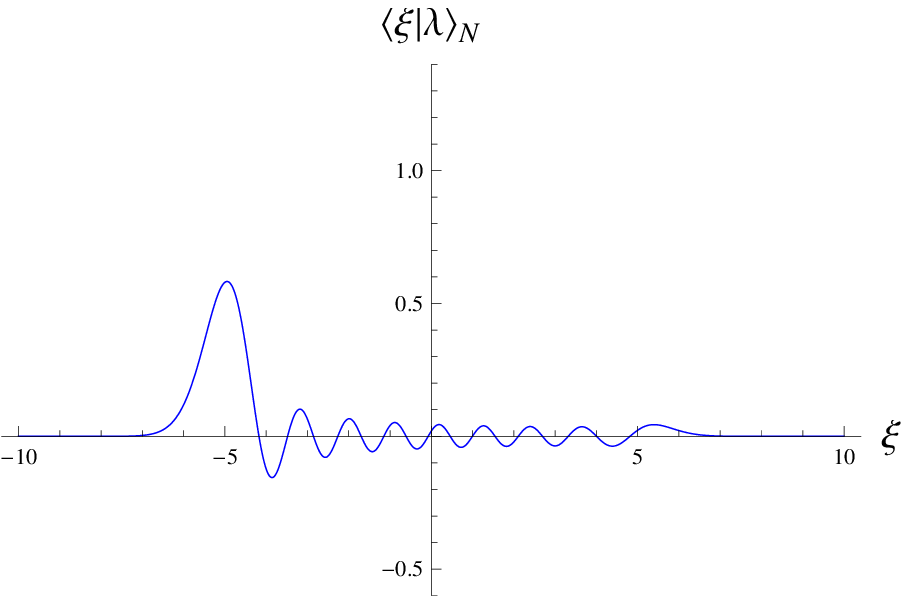}
  \label{WF0}
}
\subfigure[$\lambda=-2$]{
  \includegraphics[width=0.4\textwidth]{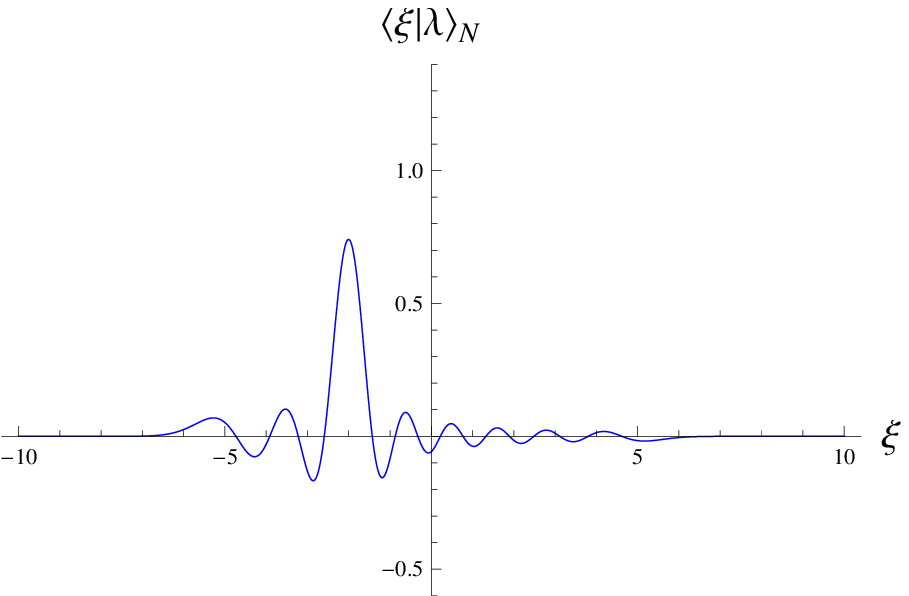}
  \label{WF2}
}
\subfigure[$\lambda=0$]{
  \includegraphics[width=0.4\textwidth]{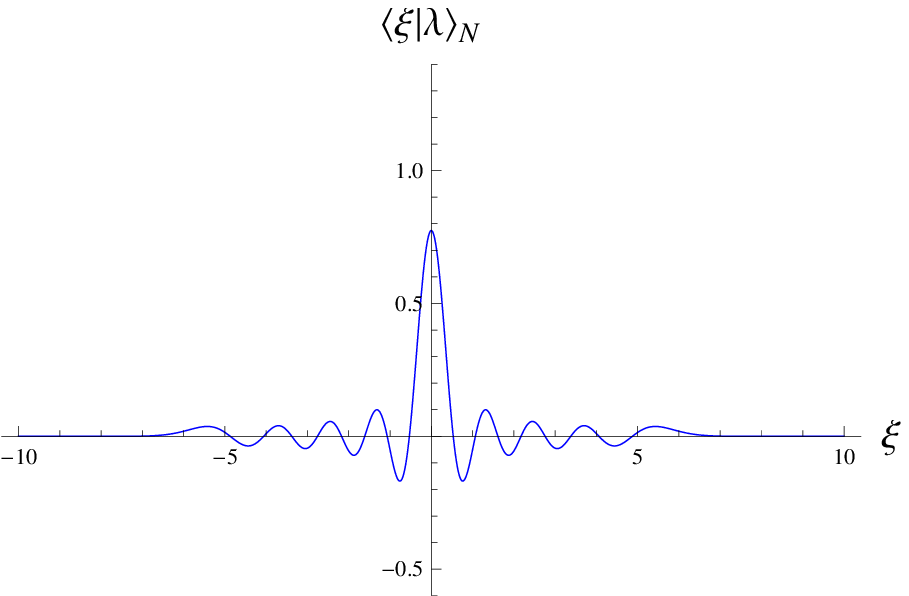}
  \label{WF4}
}
\subfigure[$\lambda=2$]{
  \includegraphics[width=0.4\textwidth]{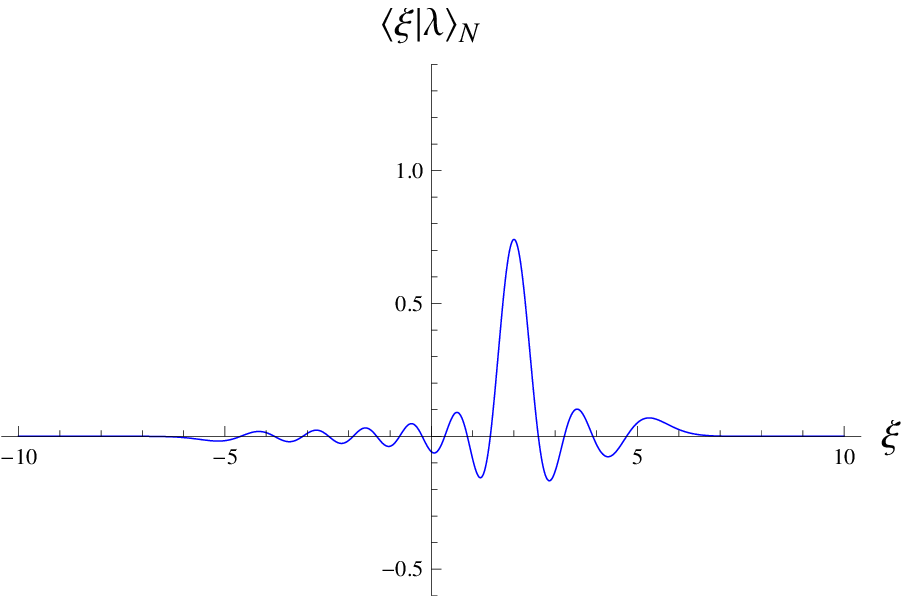}
  \label{WF6}
}
\subfigure[$\lambda=5$]{
  \includegraphics[width=0.4\textwidth]{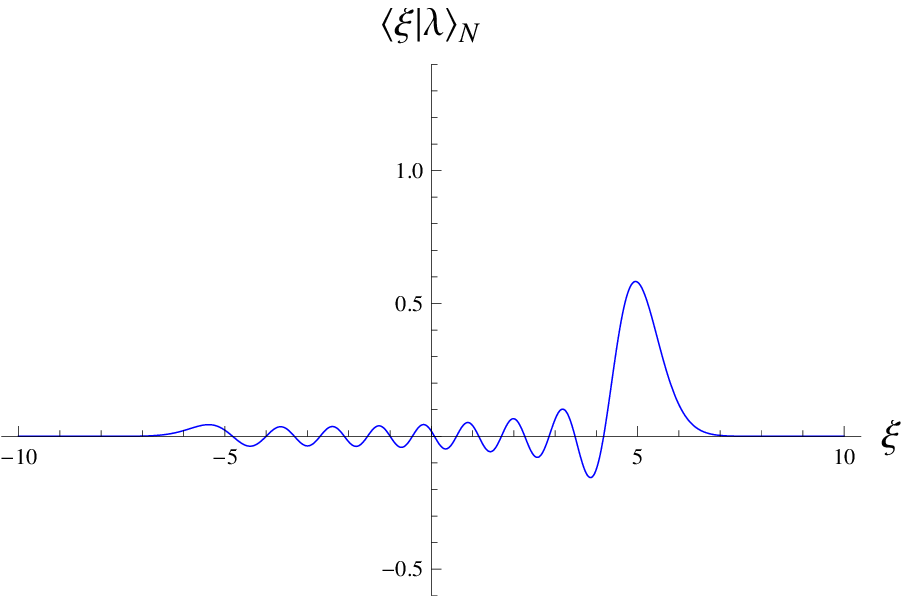}
  \label{WF8}
}
\subfigure[$\lambda=10$]{
  \includegraphics[width=0.4\textwidth]{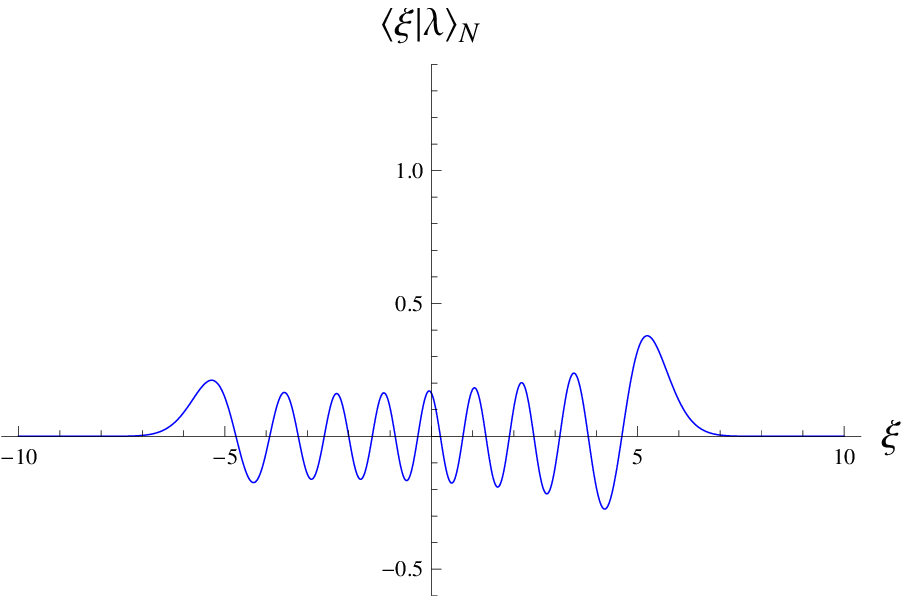}
  \label{WF10}
}
  \caption{Wavefunctions $\langle{\xi}|{\lambda}\rangle_N$ as functions of $\xi$ for $N=16$ and the indicated values of $\lambda$. }
  \label{WFs}
\end{figure}

The wavefunctions can be found, again using the Christoffel-Darboux sum formula, now in its bivariate form, \eqref{CDdoble}:
\begin{align*}
\bracket{\xi}{\lambda}_N
&=c_N(\lambda)\sum_{n=0}^N \frac{H_n(\lambda)}{\sqrt{2^n n!}}\bracket{\xi}{n}
=c_N(\lambda)\sum_{n=0}^N \frac{H_n(\lambda) H_n(\xi)}{2^n n!}\frac{e^{-\xi^2/2}}{\pi^{1/4}}\\
&=\frac{c_N(\lambda)}{2^{N+1} N!}\frac{e^{-\xi^2/2}}{\pi^{1/4}}\frac{H_{N+1}(\lambda)H_N(\xi)-H_N(\lambda)H_{N+1}(\xi)}{\lambda -\xi}.
\end{align*}
If one chooses $c_N(\lambda)$ to be real and positive then one has
\begin{equation}
\bracket{\xi}{\lambda}_N= \frac{e^{-\xi^2/2}}{\sqrt{2^{N+1}N!\sqrt{\pi}}}\frac{1}{\lambda -\xi}\frac{H_{N+1}(\lambda)H_N(\xi)-H_N(\lambda)H_{N+1}(\xi)}{\sqrt{H_N(\lambda)H_{N+1}'(\lambda)- H_N'(\lambda)H_{N+1}(\lambda))}}.
\label{wavefunctions}
\end{equation}
The graphs of selected wavefunctions are shown in figure \ref{WFs}; however, it is more instructing to see the wavefunctions change with $\lambda$ in a continuous animation (see supporting material online). 

Seeing now that the wavefunctions are highly localized, it is interesting to calculate the expected value of the position and its dispersion. There is an initial question of whether to use the truncated position, $\hxi_N$, or its full version, \hxi, but it turns out that their expected values coincide: since $\ket{\lambda}_N$ is in $\hil^N$, the range of $\hPi_N$,
\[
\tensor[_N]{\matrixel{\lambda}{\hxi}{\lambda}}{_N}=\tensor[_N]{\matrixel{\lambda}{\hPi_N\hxi\hPi_N}{\lambda}}{_N}=\tensor[_N]{\matrixel{\lambda}{\hxi_N}{\lambda}}{_N}.
\]
This is convenient, since we know already from \eqref{casieigenvector} the action of $\hxi_N$ on the vectors $\ket{\lambda}_N$. We thus have
\begin{align*}
\tensor[_N]{\matrixel{\lambda}{\hxi_N}{\lambda}}{_N}
& = \tensor[_N]{\bra{\lambda\vphantom{\hxi}}}{}\left[\lambda\ket{\lambda}_N-\frac{1}{2}\frac{c_N(\lambda)}{\sqrt{ 2^N N!}} H_{N+1}(\lambda) \ket{N}\right] \\
& =\lambda  -\frac{1}{2}\frac{c_N(\lambda)}{\sqrt{ 2^N N!}} H_{N+1}(\lambda) \cdot \!\!\tensor[_N]{\bracket{\lambda}{N}}{}\\
& =\lambda  -\frac{1}{2}\frac{|c_N(\lambda)|^2}{2^N N!} H_{N+1}(\lambda)H_N(\lambda) \\
& =\lambda  -\frac{1}{2}\frac{ H_{N+1}(\lambda)H_N(\lambda) }{(N+1)H_N(\lambda)^2-NH_{N+1}(\lambda)H_{N-1}(\lambda)}.
\end{align*}

In particular, the expected value $\langle\hxi_N\rangle$ coincides with the pseudo-eigenvalue $\lambda$ not only on the zeros of $H_{N+1}$, when it should since $\ket{\lambda}_N$ is then an eigenvector, but also on the zeros of $H_N$. This is important since these second values are interspersed with the first and therefore occur on the peaks of the oscillations of $d_N(\lambda)$, a further indication that the treatment of $\ket{\lambda}_N$ as an approximate eigenvector is appropriate.

\begin{figure}[htbp]
	\centering
		\includegraphics[width=0.5\textwidth]{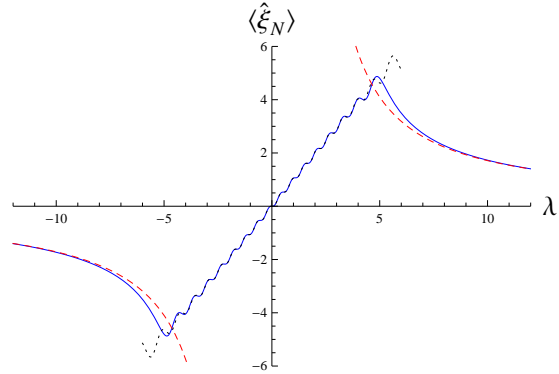}
	\caption{Expected value $\langle\hxi_N\rangle$ as a function of $\lambda$ for $N=16$, with the approximations \eqref{VEaproxcuad} and \eqref{VEaproxosc} shown dashed and dotted, respectively.}
	\label{valoresperado}
\end{figure}

Approximating the expression for the expected value of the position using the same methods that were used to approximate $d_N(\lambda)$ by expressions \eqref{aproxes}, one obtains
{\footnotesize
  \begin{subequations}
  \label{VEaproxes}
  \begin{empheq}[left=\empheqlbrace]{align}
  \langle\hxi\rangle&=\frac{N}{\lambda}+\frac{N(N-2)}{2\lambda^3}+\cdots \nonumber \\
  &\qquad\text{for} \,\lambda >\sqrt{2N+1},\,\text{and}\label{VEaproxcuad}\\
  \langle\hxi\rangle&\approx\lambda-\frac{
  \sin\left(\frac{\lambda}{\sqrt{2N+2}}    \left(1+\frac{1}{6}\frac{\lambda^2}{2N+2}+\cdots\right)\right)
  +(-1)^N\sin\left(2\sqrt{2N+2}\lambda\left(1+\frac{1}{6}\frac{\lambda^2}{2N+2}+\cdots\right)\right)
  }{\sqrt{2N+1}\left[
  1+\cos\left(\frac{2}{\sqrt{2N+1}}\lambda\left(1+\frac{1}{12}\frac{\lambda^2}{2N+1}+\cdots\right)\right)
  \right]} \nonumber \\
  &\qquad \text{for}\,\lambda<\sqrt{2N+1}.\label{VEaproxosc}
  \end{empheq}
  \end{subequations}
}
The graph of $\tensor[_N]{\matrixel{\lambda}{\hxi_N}{\lambda}}{_N}$ as a function of $\lambda$ is shown, along with both of the above aproximations (with the series truncated as shown above), in figure \ref{valoresperado}.

Since the amplitude of the oscillations in \eqref{VEaproxosc} does decrease uniformly, in the $\lambda\ll\sqrt{2N+1}$ regime one can consider $\lambda$ to be an excellent approximation to the expectation value $\langle\hxi_N\rangle$. In general, $_N\!\langle\lambda|\hxi_N|\lambda\rangle_N$ covers all of the values between the smallest and largest eigenvalue of $\hxi_N$, i.e. the first and the last zeros of $H_{N+1}$ (which is approximately the region $-\sqrt{2N+1}<\lambda<\sqrt{2N+1}$) \changed{linearly in $\lambda$ except for oscillations whose amplitude tends to zero}, although near the edges of this interval it \textit{can} occur that more than one $\lambda$ correspond to the same expectation value, which happens because the amplitude of the oscillations increases with $\lambda$.

The decay of $\langle\hxi_N\rangle$ to zero as $\lambda\rightarrow\infty$ is also interesting in its own right, since it represents the fact that in that regime the states default to the last number ket, \ket{N}, which has expected value 0 for the position.

We close this section by giving an explicit expression for the position dispersion. This is complicated by the fact that the second moments of $\hxi$ and $\hxi_N$ do \textit{not} coincide, since $\hPi_N\hxi^2\hPi_N\neq\hxi^2_N$. In fact, one has
\begin{align*}
\tensor[_N]{\matrixel{\lambda}{\hxi^2}{\lambda}}{_N} & = \tensor[_N]{\matrixel{\lambda}{\hPi_N\hxi\hPi_N^2\hxi\hPi_N}{\lambda}}{_N}+ \tensor[_N]{\matrixel{\lambda}{\hPi_N\hxi\left(1-\hPi_N\right)^2\hxi\hPi_N}{\lambda}}{_N} \\
& = \tensor[_N]{\matrixel{\lambda}{\hxi^2_N}{\lambda}}{_N}+ \frac{N+1}{2}\left|\bracket{N}{\lambda}_N\right|^2.
\end{align*}

We therefore calculate
\begin{align*}
\tensor[_N]{\matrixel{\lambda}{\hxi^2_N}{\lambda}}{_N}
& = \tensor[_N]{\bra{\lambda\vphantom{\hxi}}}{}\hxi_N\left[\lambda\ket{\lambda}_N-\frac{1}{2}\frac{c_N(\lambda)}{\sqrt{ 2^N N!}} H_{N+1}(\lambda) \ket{N}\right] \\
& = \lambda^2
-\frac{1}{2}\frac{|c_N(\lambda)|^2}{ 2^N N!} H_{N+1}(\lambda)\left(\lambda  H_{N}(\lambda) + N H_{N-1}(\lambda) \right),
\end{align*}
which gives, for the position uncertainty,
\begin{align*}
\tensor[_N]{\matrixel{\lambda}{\left(\hxi_N-\langle\hxi_N\rangle\right)^2}{\lambda}}{_N}
& = \tensor[_N]{\matrixel{\lambda}{\hxi^2_N}{\lambda}}{_N} -\tensor*[_N]{\matrixel{\lambda}{\hxi_N}{\lambda}}{_N^2} \\
& =d_N(\lambda) \left(1-\frac{|c_N(\lambda)|^2}{2^N N!}H_N(\lambda)^2 \right).
\end{align*}
\begin{figure}[htb]
	\centering
		\includegraphics[width=0.80\textwidth]{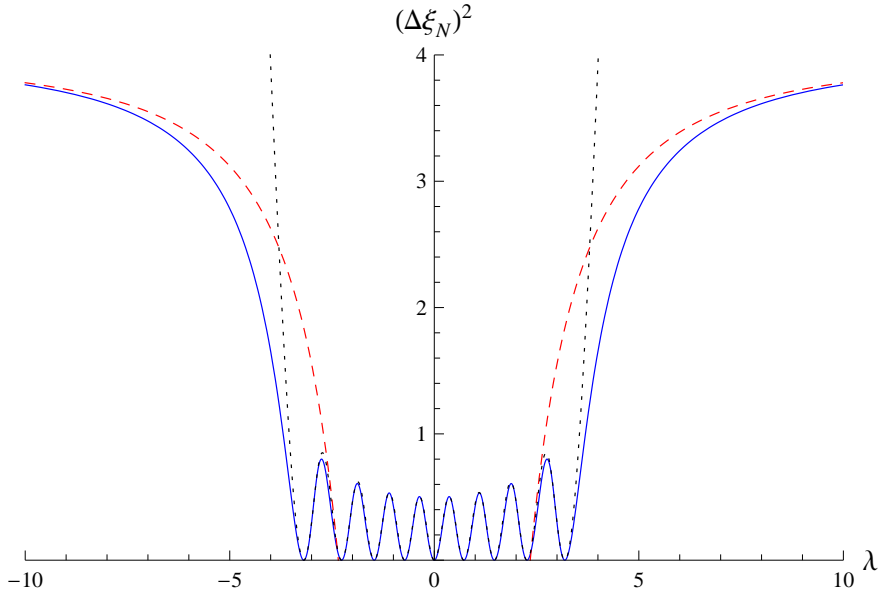}
	\caption{Position dispersion $(\Delta \hxi_N)^2$ as a function of $\lambda$ for $N=8$, along with the (dashed and dotted) approximations \eqref{dispaproxes}.}
	\label{disp_truncada}
\end{figure}
This can be approximated by the expressions
\begin{subequations}
\label{dispaproxes}
\begin{empheq}[left=\empheqlbrace]{align}
\left(\Delta \hxi_N\right)^2 & = \frac{N}{2}-\frac{N(N+3)}{4\lambda^2}+\cdots  &\qquad\text{for} \,\lambda >\sqrt{2N+1},\,\text{and}\label{disaproxcuad}\\
\left(\Delta \hxi_N\right)^2 & = d_N(\lambda)+\Or (N^{-1})   &\qquad \text{for}\,\lambda<\sqrt{2N+1},\label{dispaproxosc}
\end{empheq}
\end{subequations}
which are shown in figure \ref{disp_truncada}.

The full position dispersion, $\left(\Delta\hxi\right)^2$, is slightly more complicated. As said above, it differs in one term from the truncated-position dispersion, and can therefore be written
\begin{equation}
\tensor[_N]{\matrixel{\lambda}{\left(\hxi-\langle\hxi\rangle\right)^2}{\lambda}}{_N} = d_N(\lambda) \left(1-\frac{|c_N(\lambda)|^2}{2^N N!}H_N(\lambda)^2 \right)+\frac{N+1}{2}\frac{|c_N(\lambda)|^2}{2^N N!} H_N(\lambda)^2.
\end{equation}

\begin{figure}[htb]
	\centering
		\includegraphics[width=0.80\textwidth]{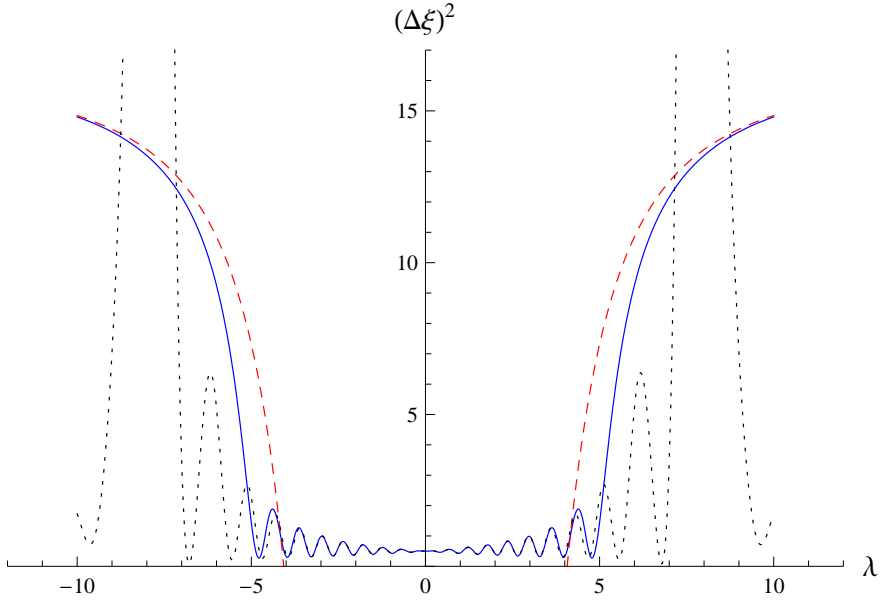}
	\caption{Dispersion of the (non-truncated) position, $(\Delta\hxi)^2$, for $N=16$, together with the approximations \eqref{dispcompaproxcuad}, dotted, and \eqref{dispcompaproxosc}, dashed. At $\lambda=0$ the dispersion is $\tfrac{1}{2}$, independently of $N$, and there the function is increasingly flatter as $N\rightarrow\infty$.}
	\label{disp_completa}
\end{figure}

Because of the $N+1$ factor, the last term does not tend to zero as $N\rightarrow\infty$ and is therefore not negligible. Furthermore, since the extra term is proportional to $H_N(\lambda)$ and the original $d_N(\lambda)$ is proportional to $H_{N+1}(\lambda)$, in the oscillatory region the two contributions are out of phase and therefore interfere constructively to make an (almost) constant function close to $\lambda=0$. As $\lambda$ grows, the differing frequencies of oscillation of both terms make them interfere and $\left(\Delta\hxi\right)^2$ does show some oscillations. This can be seen in figure \ref{disp_completa}, and is made precise by the approximations
{\footnotesize
  \begin{subequations}
  \label{dispcompaproxes}
  \begin{empheq}[left=\empheqlbrace]{align}
  (\Delta \hxi)^2 & = \frac{2N+1}{2}-\frac{N(N+2)}{2\lambda^2}-\frac{N(2N^2+2N-9)}{4\lambda^4}+\cdots \nonumber\\ 
  &\qquad\text{for} \,\lambda >\sqrt{2N+1},\,\text{and}\label{dispcompaproxcuad}\\
  (\Delta \hxi)^2 & = \frac{
  1+(-1)^N \sin\left(2\sqrt{2N+2}\lambda\left(1-\frac{1}{6}\frac{\lambda^2}{2N+2}+\cdots\right)\right)     
  \sin\left(\frac{\lambda}{\sqrt{2N+2}}\left(1+\frac{1}{6}\frac{\lambda^2}{2N+2}+\cdots\right)\right)
  }{
  1+\cos\left(\frac{2}{\sqrt{2N+1}}\lambda\left(1+\frac{1}{12}\frac{\lambda^2}{2N+1}+\cdots\right)\right)
  }
  \nonumber\\
     &\qquad \text{for}\,\lambda<\sqrt{2N+1}.\label{dispcompaproxosc}
  \end{empheq}
  \end{subequations}
}

For large $\lambda$, both dispersions are bounded, which they must since they are the expectation values of finite-dimensional operators. It is interesting to note that the full-position operator dispersion, $(\Delta\hxi)^2$, seems to be bounded from \textit{below}, in the sense that $(\Delta\xi)^2\rightarrow\tfrac{1}{2}$ uniformly in any compact set $\lambda\in[-\lambda_0, \lambda_0]$, which contrasts with the fact that $\hxi$ has eigenstates with zero dispersion. Although the latter do have infinite norm, it is well known that squeezed states with finite norm and \changed{position dispersion below the Heisenberg limit of $\tfrac{1}{2}$ exist.}

\changed{

For completeness, we include here some additional results. The inner product between different (pseudo-)eigenvectors $\ket{\lambda}_N$ is given by
\begin{equation}
\tensor[_N]{\bracket{\lambda'}{\lambda}}{_N}=\frac{c_N(\lambda')^\ast c_N(\lambda)}{2^{N+1}N!}\frac{H_{N+1}(\lambda)H_{N}(\lambda')-H_{N}(\lambda)H_{N+1}(\lambda')}{\lambda-\lambda'};
\label{innerproduct}
\end{equation}
in particular, it is zero when both $\lambda$ and $\lambda'$ are roots of $H_{N+1}$, but \textit{also} when they are roots of $H_N$, which is a consequence of the fact that they are eigenstates of $\hxi_{N-1}$, and are therefore well treated as pseudo-eigenstates of $\hxi_N$. The matrix element
\begin{equation}
	\tensor[_N]{\matrixel{\lambda'}{\hxi}{\lambda}}{_N}=\frac{c_N(\lambda')^\ast c_N(\lambda)}{2^{N+1}N!}\frac{\lambda'H_{N+1}(\lambda)H_{N}(\lambda')-\lambda H_{N}(\lambda)H_{N+1}(\lambda')}{\lambda-\lambda'}
	\label{ximatrixel}
\end{equation}
is also of interest;  the matrix element $\tensor[_N]{\matrixel{\lambda'}{\hpi}{\lambda}}{_N}$ can also be calculated.

}

\section{The Jacobi matrix}
\label{final}
In this section we briefly explore some mathematical results \changed{related to} the work described in this article. Most importantly, we retake the expression for the characteristic polynomials of the truncated position, $\hxi_N$, given in \eqref{polcarhermite}, and which can be written as
\begin{equation}
H_{N+1}(\lambda)=2^{N+1}\det(\hxi_N+\lambda)=2^{N+1} \det
\left(\begin{array}{cccccc}
\lambda &     \frac{1}{\sqrt{2}}    &     0    & \cdots & 0 & 0 \\
\frac{1}{\sqrt{2}} &     \lambda    & \frac{\sqrt{2}}{\sqrt{2}} & \cdots & 0 & 0 \\
0 & \frac{\sqrt{2}}{\sqrt{2}} &     \lambda    & \cdots & 0 & 0 \\
\vdots & \vdots & \vdots & \ddots & \vdots & \vdots \\
0 &     0    &     0    & \cdots &     \lambda    & \frac{\sqrt{N}}{\sqrt{2}} \\
0 &     0    &     0    & \cdots & \frac{\sqrt{N}}{\sqrt{2}} &    \lambda 
\end{array}\right).
\label{jacobimatrix}
\end{equation}

This can now be turned around, and viewed as a representation of the Hermite polynomials. This representation in terms of the determinant of a tridiagonal matrix exists for all families of orthogonal polynomials and is known in the special-functions literature as the Jacobi-matrix representation \cite{Askey}. Jacobi himself used this expression, as well as the explicit diagonalization given above of the tridiagonal matrix, to study Gaussian quadrature \cite{Gautschi}, that is, the approximation of the integral of a function by a finite sum of evaluation on a number of points. These points are the zeros of appropriately chosen polynomials, which can be efficiently evaluated by this method since it uses the algorithms of linear algebra.

An interesting application of this development comes if one couples the expression \eqref{jacobimatrix} with the Cayley-Hamilton theorem, which states that every matrix satisfies its own characteristic polynomial. This means that the truncated quadratures $\hxi_N$ are matrix solutions to the equation 
\begin{equation}
H_{N+1}(X)=0,
\label{ceromatricial}
\end{equation}
which is not trivial, particularly when one considers that for large $N$ the real solutions are known to exist but there is no known simple or elementary exact formula for them.

Furthermore, since each zero of $H_{N+1}$ appears exactly once as an eigenvalue of $\hxi_N$, this means that the Hermite polynomials are not only the characteristic polynomials of the matrices $\hxi_N$, but they are also their minimal polynomials (the minimal polynomial $p$ of a matrix $A$ is the \changed{unique} monic polynomial of minimal degree for which $p(A)=0$). Although there are other matrix zeros of \eqref{ceromatricial}, the condition of having the Hermite polynomials as minimal polynomial specifies the truncated quadratures $\hxi_N$ uniquely, up to matrix equivalence.

Finally, the work in this article may be used to obtain results for the Hermite polynomials which would be hard to obtain in other ways. We give one example: by numbering the zeros $\lambda_k$ of $H_{N+1}(\lambda)=0$ with $k=0,1,\ldots,N$ we obtain an othornormal basis $\{\ket{\lambda_k}_N:k=0,1,\ldots,N\}$, in terms of which the orthonormalization of the number kets reads
\[
\delta_{mn}=\bracket{m}{n}=\sum_{k=0}^N \bracket{m}{\lambda_k}_N \tensor[_N]{\bracket{\lambda_k}{n}}{} 
  = \frac{1}{ \sqrt{2^m m! 2^n n!}} \sum_{k=0}^N |c_N(\lambda_k)|^2 H_m(\lambda_k)H_n(\lambda_k),
\]
\changed{an} expression in which the sum is over all the zeros of $H_{N+1}$ and would therefore be particularly hard to obtain by ordinary means.

\section{Conclusions}
We have obtained the spectrum of the position quadrature operator $\hat\xi_N$ for a finite number $N$ of photons in terms of the zeroes of Hermite polynomials, as well as its eigenvectors in terms of lower-degree Hermite polynomials evaluated on these zeros. The normalization of the eigenstates is given through the Christoffel-Darboux kernel, and a regular structure for these is shown numerically. Approximate eigenstates for $\hat\xi_N$ were naturally defined, which represent highly localized wavefunctions centered around any value between the least and greatest zeros of $H_{N+1}$. By using an appropriate notion of limit for these spectra, in the sense that they tend to the full spectrum of the infinite-dimensional position operator $\hxi$, we showed that the finite spectrum of the truncated phase operators tends to the complete unit circle, \changed{as one would expect.} The same procedure can be followed for any arbitrary field quadrature.

\section*{Acknowledgments}
The authors wish to thank A. Frank and K. B. Wolf  for discussions. This work was partially supported \changed{by CONACYT, SEP and} by DGAPA-UNAM under project IN102811.

\end{document}